\documentclass[12pt,a4paper]{elsart}
\usepackage{latexsym,amssymb,epsfig}

\newcommand{\ea}{\mbox{\Large $\rightarrow$}}

\newcommand{\wea}{\mbox{\Large $\leftrightarrow$}}

\def\0{\phantom{0}}

\begin{document}
\pagenumbering{arabic}
\baselineskip25pt

\begin{center}
{\bf \large Chemical potential of quadrupolar 
two-centre Lennard-Jones fluids by gradual insertion} \\
\bigskip
\renewcommand{\thefootnote}{\fnsymbol{footnote}}
Jadran Vrabec\footnote[1]
{author for correspondence, tel.: +49-711/685-6107, 
fax: +49-711/685-7657, \\ email: vrabec@itt.uni-stuttgart.de}, Matthias Kettler, Hans Hasse
\renewcommand{\thefootnote}{\arabic{footnote}} \\
Institut f\"ur Technische Thermodynamik und Thermische Verfahrenstechnik, \\
Universit\"at Stuttgart, D-70550 Stuttgart, Germany 
\end{center}
 

\bigskip
\bigskip
\bigskip
\noindent
{\bf Abstract}

\noindent
The gradual insertion method for direct calculation of the chemical
potential by molecular simulation is applied in the NpT ensemble to different 
quadrupolar two-centre Lennard-Jones fluids at high density state points. 
The results agree well with Widom's test particle insertion but show 
at very high densities
significantly smaller statistical uncertainties. The gradual insertion
method, which is coupled here with preferential sampling,
extends the density range where reliable information on the chemical
potential can be obtained. Application details are reported.


\section{Introduction}
All molecular simulation techniques for the calculation of the
chemical potential which are accompanied with trial insertion (or
deletion) of real or ghost particles become inaccurate and finally
fail in the case of high densities, especially when applied to molecular 
fluids with strong interactions. This is the case for the conventional 
test particle insertion method by Widom \cite{widom6328}
as well as for the Gibbs ensemble (GE) \cite{gibbs} and grand
canonical Monte-Carlo ensemble (GC) \cite{allentilde}, which demand
insertion and deletion attempts of real particles.

Therefore, different Monte-Carlo techniques have been proposed to
improve the efficiency of the calculation of the chemical potential. 
Recent developments are, e. g., the expanded ensemble methods
\cite{voron1,nezbeda9139}, the 
particle deletion scheme \cite{boulougouris9990}, the augmented
grand canonical ensemble \cite{kaminski9449}, or the scaled particle 
Monte-Carlo method \cite{barosova9642}. An overview is given
by Kofke and Cummings \cite{kofke9797}.

In the present work, an expanded ensemble method is used. A first version 
was published by Vorontsov-Velyaminov et al. \cite{voron1}. That 
method was applied for gradual insertion by Nezbeda and Kolafa 
\cite{nezbeda9139}, who additionally introduced preferential sampling. 
Related work with expanded ensemble methods is described e.g. in
\cite{voron2,voron3,atta}.

We use the gradual insertion method as proposed by Nezbeda and Kolafa
\cite{nezbeda9139} which is based on the following idea: instead of 
inserting or deleting a real particle, a fluctuating particle is introduced, 
that undergoes changes  in a set of different discrete 'sizes' or states of 
coupling with all other particles. Additionally, the preferential sampling 
of the fluid in the vicinity of the fluctuating particle is introduced for
a further increase of accuracy.

This method was applied to different 
ensembles (NVT \cite{nezbeda9139}, GC \cite{nezbeda9139}, 
GE \cite{strnad9919}, and NpT \cite{vortler93,vortler01})
using different interaction potentials such as 
the hard sphere fluid \cite{nezbeda9139}, the square-well fluid 
\cite{strnad9919}, the spherical Lennard-Jones fluid 
\cite{vortler93,kolafa9330}, and to ternary mixtures of hard spheres 
and hard heteronuclear diatomics \cite{strnad}.

In this work we apply the gradual insertion method for the first time
to a polar intermolecular potential, the two-centre Lennard-Jones plus
point quadrupole fluid. That potential allows to describe the thermodynamic
properties of many real fluids, like carbon dioxide, ethyne, propyne, 
or carbon disulfide with good accuracy, as has been shown in a comprehensive 
study \cite{vrabec2001,stoll2001}. The present study is performed in 
the NpT ensemble at high densities. The results are critically compared to
test particle insertion. Details on the gradual insertion method, which 
is used here together with preferential sampling, are given.

\section{Method}
In the following we briefly summarize the gradual insertion method in a
NVT ensemble, for more details see \cite{nezbeda9139}. 
The key element of the method is to introduce a 'fluctuating 
particle' which can appear at different states of coupling with all 
other particles. This set of states starts from a pointwise, completely 
interactionless or decoupled state, up to the fully coupled state, 
in which the fluctuating particle has the same properties as the other 
particles. Each of the partially coupled states is related to a NVT 
sub-ensemble with no physical meaning. These sub-ensembles can be 
depicted in a scheme as follows:
\begin{equation}
\begin{array}{ccccccccc}
[N] & \wea & [N+\psi_1] & \wea & \cdots & \wea & [N+\psi_{k-1}] & \wea & [N+\psi_{k}], \\
    w_0 && w_1 &&&& w_{k-1} && w_{k}
\end{array}
\label{nksch}
\end{equation}
where $[N+\psi_j]$ denotes a sub-ensemble with $N$ regular particles
and one fluctuating particle in the state $j$, with $j=0,\dots,k$.  The
interaction energy of the fluctuating particle with the remaining
particles is denoted by $\psi_j$, in the fully coupled state by
$\psi_k$, and in the fully decoupled state it is $\psi_0=0$.
Note that $[N+\psi_k]$ equals $[N+1]$.
The arrows describe the transitions between neighbouring states.
Finally, $w_j$ denotes the weight of the state $j$.

Additionally to the standard Monte-Carlo moves in a NVT ensemble,
trial translation and trial rotation, the trial change of fluctuating 
particle state is necessary. The probability of accepting a change of 
the fluctuating particle from state $i$ to
state $j$ is given by
\begin{equation}
P_{acc}(i \ea j) = {\rm min} \left(1,
\frac{w_j \cdot \Pi(j\ea i)}{w_i \cdot \Pi(i\ea j)} \cdot
\exp\{-\beta(\psi_j-\psi_i)\}\right),
\end{equation}
where $\Pi(j\ea i)$ and $\Pi(i\ea j)$ denote a priori transition probabilities 
of the respective changes and $\beta$ is the Boltzmann factor. 

The chemical potential in the NVT ensemble is obtained from
\cite{nezbeda9139}
\begin{equation}
\mu = \mu^{id}(T) + kT \cdot \ln \frac{N}{V} + kT \cdot 
\ln\left\langle \frac{w_{k}}{w_0} \cdot \frac{ {\rm Prob}[N]} {{\rm Prob}[N+1]}
\right\rangle,
\end{equation}
where ${\rm Prob}[N]$ and ${\rm Prob}[N+1]$ are the probabilities
to observe an ensemble with $N$ and $N+1$ particles, respectively.
The paranthesis $<>$ denote the ensemble average.
Furthermore, $\mu^{id}(T)$ is the temperature dependent ideal 
part of the chemical potential. If the fluctuating particle is in the 
fully decoupled state, an insertion attempt into a random new position 
follows. When the fluctuating particle reaches the fully coupled state 
another particle is chosen randomly, to be treated as the fluctuating one.

Improvement of efficiency often can be obtained by using the
preferential sampling method, see e. g. Allen and Tildesley
\cite{allentilde}. The idea is to sample particles preferentially in the 
vicinity of the fluctuating particle. Therefore, a function $f(r)$ is 
introduced, denoting the probability of an attempt to move a host particle 
in a distance $r$ from the fluctuating particle.  The following steps are
necessary to keep the microreversibility condition \cite{nezbeda9139}:
\begin{itemize}
\item[(i)] random choice of a host particle, 
\item[(ii)] acceptance of this choice with the probability
$f(r_{old})$, 
\item[(iii)] random choice of a new position and orientation of
the host particle, 
\item[(iv)] acceptance of this choice with the probability 
${\rm min}(1,f(r_{new}) /f(r_{old}))$, 
\item[(v)] the new configuration is accepted with the probability 
\mbox{${\rm min}(1,\exp\{-\beta(U_{new}-U_{old})\})$},
\end{itemize}
where $U$ denotes the configurational energy.

In order to extend the gradual insertion method to the NpT 
ensemble the volume fluctuation step has to be considered additionally. 
To set the pressure $p$, a volume change from $V_{old}$ to $V_{new}$
is accepted with the probability \cite{allentilde}
\begin{equation}
\!\!\!\!\!\!\!\!\!P_{acc} (V_{old} \ea V_{new}) = \min \left( 1, 
(V_{new}/V_{old})^N \cdot\exp \left\{ 
-\beta (p (V_{new}-V_{old})+ U_{new} - U_{old} ) \right\} \right).
\end{equation}
This step was applied at random position within the Markov chain,  
independent from the state of the fluctuating particle.
Similar to the NVT ensemble, only configurations with a fully coupled
fluctuating particle contribute to the Monte-Carlo sampling of
physical properties.

As the volume fluctuates in the NpT ensemble, Eq. (3) has to be modified
to claculate the chemical potential
\begin{equation}
\mu = \mu^{id}(T) + kT \cdot \ln\left\langle \frac{N}{V} \cdot \frac{w_{k}}{w_0}
\cdot \frac{{\rm Prob}[N]}{{\rm Prob}[N+1]} \right\rangle.
\end{equation}
Up to our knowledge, no formal derivation of Eq. (5) has been given in the
literature so far. It is not within the scope of this work 
to close that gap. We will limit ourselves to show that the results
from Eq. (5) are in agreement with Widom's method, cf. section 4.

\section{Investigated model and technical details}
In the present investigation we consider the two-centre Lennard-Jones
plus pointquadru\-pole fluid (2CLJQ). It is composed of two identical
Lennard-Jones sites a distance $L$ apart plus an ideal pointquadrupole
of moment $Q$ placed in the geometric centre of the molecule \cite{gray84}. 
The charges of the quadrupole are arranged along the molecular axis in the
symmetric sequence $+$, $-$, $-$, $+$ or, having the same energetic
effect in pure fluids, $-$, $+$, $+$, $-$. A detailed description of this
fluid is given in \cite{stoll}. 

The Lennard-Jones parameters $\sigma$ and $\epsilon$ of the 2CLJQ pair 
potential were used for the reduction of the thermodynamic properties 
and the model parameters $L$ and $Q$: 
temperature $T^*=kT/\epsilon$, pressure $p^*=p\sigma^3/\epsilon$, density 
$\rho^*=\rho\sigma^3$, configurational energy $u^*=u/\epsilon$, 
elongation $L^*=L/\sigma$, squared quadrupolar moment 
$Q^{*2}=Q^2/\left( \epsilon\sigma^5 \right)$.

For the simulation runs $N=512$ particles were used, the cut-off
radius was set to $4~\sigma$, applying periodic boundary conditions and the
minimum image convention.
The long range corrections for the two-centre Lennard-Jones potential 
\cite{lustig8817} were considered, in the case of quadrupolar interactions 
no long range corrections have to be made.The number of loops, defined 
below, was 50 000. The maximum values of translation distance, rotation 
angle, and volume change were adjusted to yield acceptance rates of 
roughly 0.5. Statistical uncertainties were calculated by 
conventional block averaging \cite{fincham8645}.

The number of states of the fluctuating particle was chosen to $k=6$. 
For the fluctuating particle in the state $j$ we set 
$\sigma_j/\sigma=(1/2+\sqrt{j/24})$, $\epsilon_j/\epsilon=j/k$, $L_j/L=1$,
and $Q^2_j/Q^2=j/k$. The diameter of the hard sphere, shielding the 
quadrupolar interaction site, was set to $0.4~\sigma$, except for the 
fully decoupled state $j=0$, where it is zero. The shielding does not allow 
quadrupolar interaction sites to approach closer to each other than 
$0.4~\sigma$. Finally, the weights $w_j$ had to be chosen. Starting from unity
for all states, they were adjusted during an equilibration period in order 
to achieve a roughly equal distribution in the occurrence of all states.

One Monte-Carlo loop is defined here as $N$ trial translations, $2/3\cdot N$
trial rotations, and 1 trial volume change, which are the regular NpT moves.
The additional gradual insertion moves within a loop are: $M_C$$\cdot N$ attempts
to change the state of the fluctuating particle, $M_F$$\cdot N$ attempts to 
translate the fluctuating particle, $2/3\cdot$$M_F$$\cdot N$ attempts to 
rotate the fluctuating particle, $M_P$$\cdot N$ attempts to find a host particle 
for preferential translation, and $2/3\cdot$$M_P$$\cdot N$ attempts to find a 
host particle for preferential rotation. Here, $M_C$, $M_F$, and $M_P$ are the
parameters of the gradual insertion method.

\section{Results and discussion} 
In Fig. 1 running averages of the chemical potential for the 2CLJQ fluid 
($L^*=0.8$, $Q^{*2}=4$) in the liquid phase at $T/T_c\approx 0.55$
are shown as a function of Monte-Carlo loops, or respectively, molecular
dynamics time steps. The technical details of the molecular dynamics simulations
are given in \cite{stoll}.
The Monte-Carlo loops, as defined here, and the
molecular dynamics time steps should be roughly comparable in the 
sense that both result in a new configuration of the entire system.
The better convergence of gradual insertion at this high density 
state point can clearly be seen. The large steps in the Widom curves 
are typical when this method is applied to very dense fluids. They are due 
to the bad statistics, which comes from the small number of test particles 
which contribute significantly to the average. 

In Table 1, a comparison is given between simulation data from the present
work and results from Widom's method \cite{widom6328} taken from \cite{stoll}.
Six different 2CLJQ fluids are investigated 
in the liquid state close to the bubble line at two temperatures 
($T/T_c\approx 0.55$ and $T/T_c\approx 0.8$, respectively).
The first finding is, that the results of both methods agree within 
their statistical uncertainties, which is a numerical proof of Eq. (5). 
Regarding, secondly, the statistical uncertainties 
for all 2CLJQ fluids at the higher temperatures 
($T/T_c\approx 0.8$), where intermediate liquid densities are found,
both methods yield results with approximately the same statistics. 
But for $T/T_c\approx 0.55$, where the densities are roughly 20\% higher,
the uncertainties of Widoms's method increase by a factor of 10 to 15.
On the other hand, the 
uncertainties of the gradual insertion increase only by a factor of about 
2 to up to 4. So, at state points with high densities ($T/T_c\approx 0.55$) 
gradual insertion yields results that have considerably better statistics. 
For the 2CLJQ fluid with the highest elongation and highest quadrupole 
($L^*=0.8$, $Q^{*2}=4$), at the low temperature, Widom's method is close 
to failure (the uncertainty in absolute numbers is almost $\pm 1$). 
Gradual insertion gives a decent result with $\pm 0.02$.
In such very dense and srongly interacting fluids, the accuracy of 
Widom's method can not be improved significantly by using more test particles 
or by increasing the length of the simulation run.

In order to compare the computational effort, we performed 
Monte-Carlo simulations in the NpT ensemble on a conventional workstation 
applying both methods and keeping all other parameters, such as number of
particles or cut-off radius constant. It turns out, that the simulation run 
with gradual insertion ($M_C$=10, $M_F$=10, $M_P$=50) needs 19.7 CPU h and the run 
with Widom's method (2 000 test particles after each loop) needs 3.7 CPU h.
Therein, regular NpT configuration generation consumes 1.6 CPU h.
So the computational effort for gradual insertion is an order of magnitude
higher. For a comparison with the same CPU time spent for both methods, 
it would have been possible to increase the number of configurations or the
number of test particles in Widoms method by a factor of 10. Increasing the
number of test particles would have lead to no improvements, as their number is
already very high. Increasing the number of configurations by a factor of 10
would have lead to a reduction of the statistical uncertainties by about a
factor of 3 at best.

For the 2CLJQ fluid ($L^*=0.8$, $Q^{*2}=4$) at the lower temperature 
($T/T_c\approx 0.55$) a study of the influence of the free parameters 
of the gradual insertion method on the chemical potential is given in 
Table 2. The reference point is $M_C$=10, $M_F$=10, and $M_P$=50. 
These parameters were altered up and down by a factor of 2.5. 
The increase of the number of attempts to change the state of the 
fluctuating particle from $M_C$=4 to $M_C$=10 yields lower uncertainties, 
whereas upon further increase to $M_C$=25 no benefit was observed.
The variation of the number of attempts to translate or rotate the 
fluctuating particle ($M_F$) seems to have no influence on the uncertainty in 
the investigated range of that parameter. 
The number of attempts to find a host particle for preferential 
translation or rotation ($M_P$) has the clearest influence 
on the uncertainty. Note that an increase above $M_P$$\approx$100 is not 
recommended. For instance, at $M_P$=125 we have a relation of 125 preferential
translation moves to one ordinary translation move. This extreme ratio leads to
the slight deviation shown in the last line of Table 2.

\section{Conclusion} 
This investigation shows that NpT simulations with gradual insertion 
combined with preferential sampling can yield results for the
chemical potential of realistic strongly interacting molecular fluids 
at high densities with clearly improved statistics. It can be applied at state 
points where conventional test particle insertion breaks down.

\medskip
\bigskip
{\bf Acknowledgements}

\bigskip
We gratefully acknowledge financial support by Deutsche Forschungsgemeinschaft, 
Sonderforschungsbereich 412, University of Stuttgart.


\clearpage

\begin{table}[t]
\noindent
\caption{Density, configurational energy, and chemical potential for different 
2CLJQ fluids close to their bubble lines at $T/T_c \approx 0.55$ 
(upper blocks) and $T/T_c \approx 0.8$ (lower blocks). 
Method A: gradual insertion with the parameters 
$M_C$~=~10, $M_F$~=~10, and $M_P$~=~50; Method B: Widom's test particle
method \cite{stoll}.
The numbers in paranthesis denote the uncertainty in the
last digits.}
\label{1}
\bigskip
\begin{center}
\begin{tabular}{|l|l||l|l|l|c|}\hline
~~~$T^*$ & ~$p^*$ & ~~~~~~$\rho^*$ & ~~~~~~$u^{*}$ & ~~~~$\mu/kT$ & Method              \\ \hline
\multicolumn{5}{l}{$L^*=0.4$, $Q^{*2}=0$}                                               \\ \hline 
1.7875 & 0    & 0.5960 (3) &           -15.524 (8)           & -6.31\phantom{0} (2) & A \\ \cline{3-6}
       &      & 0.5952 (2) &           -15.504 (6)           & -6.34\phantom{0} (6) & B \\ \hline
2.6    & 0.06 & 0.4747 (8) &           -11.80\phantom{0} (2) & -3.867 (4)           & A \\ \cline{3-6}
       &      & 0.4718 (6) &           -11.74\phantom{0} (1) & -3.882 (7)           & B \\ \hline
\multicolumn{5}{l}{$L^*=0.4$, $Q^{*2}=4$}                                               \\ \hline 
2.09   & 0    & 0.6566 (3) &           -23.58\phantom{0} (1) & -7.05\phantom{0} (3) & A \\ \cline{3-6}
       &      & 0.6562 (2) &           -23.57\phantom{0} (1) & -6.8\phantom{00} (5) & B \\ \hline
3.04   & 0.07 & 0.5180 (8) &           -16.70\phantom{0} (3) & -3.95\phantom{0} (1) & A \\ \cline{3-6}
       &      & 0.5154 (4) &           -16.61\phantom{0} (2) & -3.98\phantom{0} (1) & B \\ \hline
\multicolumn{5}{l}{$L^*=0.6$, $Q^{*2}=0$}                                               \\ \hline 
1.4025 & 0    & 0.5008 (2) &           -12.706 (6)           & -6.59\phantom{0} (1) & A \\ \cline{3-6}
       &      & 0.5001 (2) &           -12.689 (5)           & -6.57\phantom{0} (7) & B \\ \hline
2.04   & 0.04 & 0.3947 (7) & \phantom{0}-9.51\phantom{0} (2) & -4.03\phantom{0} (1) & A \\ \cline{3-6}
       &      & 0.3931 (4) & \phantom{0}-9.47\phantom{0} (1) & -4.058 (8)           & B \\ \hline
\multicolumn{5}{l}{$L^*=0.6$, $Q^{*2}=4$}                                               \\ \hline 
1.628  & 0    & 0.5521 (2) &           -19.06\phantom{0} (1) & -7.31\phantom{0} (3) & A \\ \cline{3-6}
       &      & 0.5513 (2) &           -19.03\phantom{0} (1) & -7.6\phantom{00} (4) & B \\ \hline
2.368  & 0.04 & 0.4310 (7) &           -13.33\phantom{0} (2) & -4.16\phantom{0} (1) & A \\ \cline{3-6}
       &      & 0.4298 (5) &           -13.29\phantom{0} (2) & -4.16\phantom{0} (1) & B \\ \hline

\end{tabular}
\end{center}
\end{table}

\clearpage
\noindent
{\small Table 1 -- continued}

\bigskip
\begin{table}[ht]
\begin{center}
\begin{tabular}{|l|l||l|l|l|c|}\hline
~~~$T^*$ & ~$p^*$ & ~~~~~~$\rho^*$ & ~~~~~~$u^{*}$ & ~~~~$\mu/kT$ & Method              \\ \hline
\multicolumn{5}{l}{$L^*=0.8$, $Q^{*2}=0$}                                               \\ \hline 
1.177  & 0    & 0.4389 (2) &           -11.124 (4)           & -6.82\phantom{0} (1) & A \\ \cline{3-6}
       &      & 0.4384 (1) &           -11.109 (4)           & -6.7\phantom{00} (1) & B \\ \hline
1.712  & 0.03 & 0.3421 (6) & \phantom{0}-8.19\phantom{0} (1) & -4.161 (8)           & A \\ \cline{3-6}
       &      & 0.3398 (5) & \phantom{0}-8.14\phantom{0} (1) & -4.192 (8)           & B \\ \hline
\multicolumn{5}{l}{$L^*=0.8$, $Q^{*2}=4$}                                               \\ \hline 
1.342  & 0    & 0.4963 (2) &           -17.31\phantom{0} (1) & -7.91\phantom{0} (2) & A \\ \cline{3-6}
       &      & 0.4955 (2) &           -17.276 (8)           & -8.3\phantom{00} (8) & B \\ \hline
1.952  & 0.03 & 0.3932 (4) &           -12.36\phantom{0} (1) & -4.495 (8)           & A \\ \cline{3-6}
       &      & 0.3928 (4) &           -12.35\phantom{0} (1) & -4.51\phantom{0} (2) & B \\ \hline
\end{tabular}
\end{center}

\end{table}
   
\begin{table}[t]
\noindent
\caption{Influence of the gradual insertion parameters 
on the chemical potential of the 2CLJQ fluid ($L^*=0.8$, $Q^{*2}=4$)
at $T^*=1.342$, $p^*=0$. The numbers in paranthesis denote the uncertainty 
in the last digits.}
\label{2}
\bigskip
\begin{center}
\begin{tabular}{|r|r|r||l|}\hline
$M_C$  & $M_F$  & $M_P$  & ~~$\mu/kT$ \\ \hline
 10 &  10 &  50 & -7.97 (2)    \\ \hline
  4 &  10 &  50 & -8.00 (5)    \\ \hline
 25 &  10 &  50 & -8.04 (3)    \\ \hline
 10 &   4 &  50 & -8.00 (4)    \\ \hline
 10 &  25 &  50 & -7.99 (4)    \\ \hline
 10 &  10 &  20 & -7.97 (6)    \\ \hline
 10 &  10 & 125 & -7.94 (1)    \\ \hline
\end{tabular}
\end{center}
\end{table}

\clearpage

\listoffigures
\clearpage

\begin{figure}[ht]
\caption[Running averages of the chemical potential over Monte-Carlo 
loops, or respectively, molecular dynamics time steps of independent simulation runs for the 
2CLJQ fluid ($L^*=0.6$, $Q^{*2}=4$) at $T^*=1.628$, $p^*=0$ 
using $N=512$ particles and a cut-off radius $r_c=4~\sigma$. 
Thick line: Monte-Carlo simulation using gradual insertion with the parameters 
$M_C$=10, $M_F$=10, and $M_P$=50. 
Thin lines: Three different molecular dynamics simulation runs using Widom's 
method with 2 000 test particles at each time step.
Dashed lines: straight horizontal guide to the eye.
]{}

\label{xpic}
\begin{center}
\includegraphics[width=165mm,height=125mm]{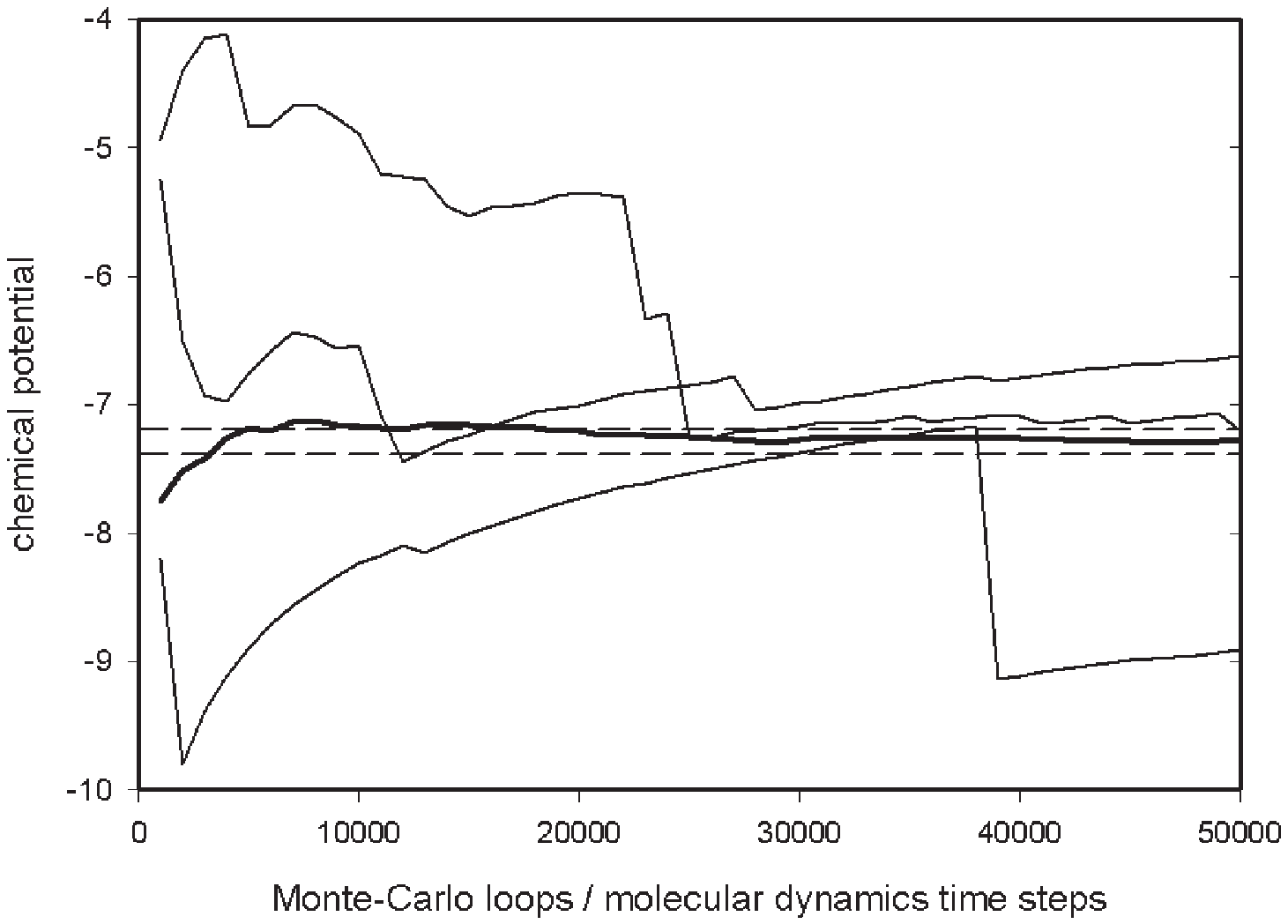}
\end{center}
\end{figure}

\clearpage


\begin{thebibliography}{99}
\bibitem{widom6328} B. Widom, J. Chem. Phys. 39 (1963) 2808.
\bibitem{gibbs} A. Z. Panagiotopoulos, Mol. Phys. 61 (1987) 813.
\bibitem{allentilde} M. P. Allen and D. J. Tildesley, Computer simulations of liquids, Clarendon Press, Oxford, 1987.
\bibitem{voron1} S. V. Shevkunov, A. A. Martinovski, and P. N. Vorontsov-Velyaminov, High Temp. Phys. (USSR) 26 (1988) 246.
\bibitem{nezbeda9139} I. Nezbeda and J. Kolafa, Mol. Simul. 5 (1991) 391.
\bibitem{boulougouris9990} G. C. Boulougouris, I. G. Economou and D. N. Theodrou, Mol. Phys. 96 (1999) 905.
\bibitem{kaminski9449} R. D. Kaminski, J. Chem. Phys. 101 (1994) 4986.
\bibitem{barosova9642} M. Barosova, A. Malievski, S. Labik, W. R. Smith, Mol. Phys. 87 (1996) 423.
\bibitem{kofke9797} D. A. Kofke, P. T. Cummings, Mol. Phys. 92 (1997) 973.
\bibitem{voron2} A. P. Lyubartsev, A. A. Martinovski, S. V. Shevkunov, and P. N. Vorontsov-Velyaminov, J. Chem. Phys. 96 (1992) 1776.
\bibitem{voron3} A. P. Lyubartsev, A. Laaksonen, and P. N. Vorontsov-Velyaminov, Mol. Simul. 18 (1996) 43.
\bibitem{atta} P. Attard, J. Chem. Phys. 98 (1993) 2225.
\bibitem{strnad9919} M. Strnad and I. Nezbeda, Mol. Simul. 22 (1999) 193.
\bibitem{vortler93} H. L. V\"ortler, Verhandl. der Dt. Phys. Ges. 17A (1993) 965.
\bibitem{vortler01} H. L. V\"ortler, Habilitationsschrift, Universit\"{a}t Leipzig, 2001.
\bibitem{kolafa9330} J. Kolafa, H.~L. V\"ortler, K. Aim and I. Nezbeda, Mol. Simul. 115 (1993) 305.
\bibitem{strnad} M. Strnad and I. Nezbeda, Mol. Phys. 85 (1995) 91.
\bibitem{vrabec2001} J. Vrabec, J. Stoll, and H. Hasse, J. Phys. Chem. B 105 (2001) 12126.
\bibitem{stoll2001} J. Stoll, J. Vrabec, and H. Hasse, AIChE J., submitted (2001).
\bibitem{gray84} C. G. Gray and K. E. Gubbins, Theory of molecular fluids, Volume 1: Fundamentals, Clarendon Press, Oxford, 1984, p. 76.
\bibitem{stoll} J. Stoll, J. Vrabec, H. Hasse, and J. Fischer, Fluid Phase Equilibria 179 (2001) 339.
\bibitem{lustig8817} R. Lustig, Mol. Phys. 65 (1988) 175.
\bibitem{fincham8645} D. Fincham, N. Quirke, and D. J. Tildesley, J. Chem. Phys. 84 (1986) 4535.
\end{thebibliography}
\end{document}